# LLM-Supported Content Analysis of Motivated Reasoning on Climate Change

**Kim, Yuheun**     Syracuse University, USA | ykim72@syr.edu
**Liu, Qiaoyi**     Syracuse University, USA | qliu11@syr.edu
**Hemsley, Jeff**   Syracuse University, USA | jjhemsle@syr.edu

**ABSTRACT**

Public discourse around climate change remains polarized despite scientific consensus on anthropogenic climate change (ACC). This study examines how "believers" and "skeptics" of ACC differ in their YouTube comment discourse. We analyzed 44,989 comments from 30 videos using a large language model (LLM) as a qualitative annotator, identifying ten distinct topics. These annotations were combined with social network analysis to examine engagement patterns. A linear mixed-effects model showed that comments about government policy and natural cycles generated significantly lower interaction compared to misinformation, suggesting these topics are ideologically settled points within communities. These patterns reflect motivated reasoning, where users selectively engage with content that aligns with their identity and beliefs. Our findings highlight the utility of LLMs for large-scale qualitative analysis and highlight how climate discourse is shaped not only by content, but by underlying cognitive and ideological motivations.

**INTRODUCTION**

The debate over climate change continues to polarize public opinion, despite the overwhelming scientific consensus affirming anthropogenic climate change (ACC). Social media platforms are key sites where these discourses unfold, enabling users to express support, doubt, or denial interactively. YouTube is especially notable, as users deliberately seek content and engage through comments.

While the academic community has long studied climate change denialism through psychological, communication, and environmental science lenses, there remains a gap in large-scale, qualitative examinations of public opinion—particularly those shaped by complex cognitive, social, and ideological motivations. This paper investigates climate change discourse in YouTube comments, focusing on both videos that affirm ("believers") and deny ("skeptics") ACC. In this study, we define "believers" as individuals who accept the scientific consensus that climate change is occurring and is the result of human activities. Climate change "skeptics" are individuals who are doubtful or actively dismissive of the prevailing scientific consensus on global warming. Our inquiry is grounded in the literature of motivated reasoning, which proposes that people's beliefs are shaped more by desires, identity, and ideological commitments than facts. Prior studies suggest that climate change denial may be driven by a desire to protect one's social group identity, avoid cognitive dissonance, or align with political ideologies. Furthermore, cognitive limitations, such as low science literacy or reliance on heuristics, also contribute to resistance against ACC. However, many of these studies have been limited to small-scale qualitative interviews or large-scale quantitative analysis with minimal qualitative depth.

By leveraging LLMs to scale up qualitative analysis, our research contributes a novel methodological approach in understanding climate change discourse online. We aim to use LLM as a qualitative annotator that extracts the underlying topics in these discussions. These topics are then analyzed in combination with the social network structures among commenters, offering a layered understanding of how climate change opinions circulate and solidify online. To guide this inquiry, we ask:

RQ: How do climate change "believers" and "skeptics" differ in their YouTube comment discourse, in respect to the topics they emphasize and the way they engage with other users and what might these differences reveal about underlying motivated reasoning?

This study contributes to research on climate change discourse and computational methods in three key ways. First, we demonstrate how large language models can be effectively employed to scale qualitative content analysis, maintaining interpretability through rationale-based prompting and cross-validation. Second, we offer a comparative analysis of how believers and skeptics structure their discourse on YouTube, showing how different topics elicit distinct engagement patterns that reflect motivated reasoning. Finally, by integrating topic modeling with social network analysis, we present a layered approach to understanding the relationship between discourse content and user interaction dynamics, revealing how ideologically aligned content circulates within fragmented communities.

## LITERATURE REVIEW
**Motivated Reasoning and Rejection of Science**

Scholars from various domains have studied the climate change denialism phenomena using different theories, with motivation reasoning as the most common and plausible explanation. Motivated reasoning refers to cognitive processes (e.g. desire or preference) that influence one's reasoning such as forming beliefs and making decisions (Kunda, 1990). Studies show climate change denial may be driven by the desire to (1) win arguments (Groenendyk & Krupnikov, 2021); (2) maintain a social group identity, or (3) seek ideological consistency (Bayes & Druckman, 2021). Other contributing factors include low science literacy, lack of cognition, and social and political identity constraints (Chen et al., 2021; Dunlap, 2013; Fischer et al., 2022; Hutmacher et al., 2024; Kovaka, 2021).

While motivated reasoning plays a crucial role in facilitating logical decisions, it can also lead to hasty and biased reasonings. When motivated for accuracy, people rely less on stereotypes and anchoring bias (Freund et al., 1985). Still, individuals have the tendency to draw desired conclusions using selective evidence, in other words, maintaining an "illusion of objectivity" (Kunda, 1990). This tendency becomes especially problematic in organized science denial, where misleading information can alter people's attitudes towards climate science (Lewandowsky & Oberauer, 2016).

Despite strong evidence for ACC, denialism continues to gain popularity, supported by contrarian scientists (Liu et al., 2025), fossil fuels corporations, conservative think tanks, and various front groups (E. Dunlap & McCright, 2011). In such context, people are prone to make 'easy' judgments using cognitive heuristics, or 'rules of thumb' to avoid confronting the 'inconvenient truth', rather than systematically review evidence on complex social issues (Hornsey et al., 2016). This could also be explained by the idea that accepting climate change equals to admitting human-responsible catastrophe, thus people respond with negative emotions to intuitively avoid that recognition (Sinatra & Hofer, 2021).

Understanding what drives motivated reasoning is key to developing more effective climate communication strategies and neutralizing polarized opinions (Bayes & Druckman, 2021). However, most research relies on small-scale population and lacks interpretive nuances in assessing motivation-driven voices. For example, Hutmacher et al. (2024) conducted a study on the influence of numeracy and cognition in motivated reasoning, but only focused on people's attitudes towards regulations on reducing $CO_2$ emissions. Meanwhile, Hornsey et al. (2016) obtained data from large research organizations and government agencies and conducted a meta-analysis of 27 variables correlated with climate change attitudes. Although these studies provide valuable insight, they fall short of addressing how motivated reasoning manifests in public discourse. Our study helps fill that gap through a thorough qualitative analysis on the factors of motivated reasoning affecting a larger population.

Recent research on climate change discourse has increasingly emphasized the role of social media platforms in intensifying polarization and shaping public understanding. While earlier work often focused on climate change denial as outright rejection of scientific consensus, more recent studies suggest that denial has evolved into more nuanced rhetorical forms. For instance, research finds that many YouTube influencers now shift from denying climate change to attacking policy measures and climate activism, often using culture war narratives that resonate with ideological identities (de Nadal, 2024). This aligns with our observation that topics like "media portrayal" or "government policy" elicit different levels of engagement depending on their alignment with group values.

These rhetorical shifts are also embedded in broader sociopolitical and networked environments. (Falkenberg et al., 2022) show that climate discourse polarization on Twitter surged during COP26, fueled by spikes in coordinated contrarian messaging. Meanwhile, other research highlight how climate beliefs often reflect deeper ideological identities, with skepticism clustering alongside right-wing populism and libertarian values (Bliuc et al., 2015; Yantseva, 2024). Such findings reinforce the relevance of motivated reasoning, as users engage with content not merely to deliberate but to affirm group norms. Furthermore, (Sultana et al., 2024; van Eck, 2024) point out that online climate discourse is increasingly characterized by emotionally charged, segregated interactions, which help explain why certain topics function as echo chamber statements in our data.

**Content Analysis with LLMs**

Content analysis is a method to attain condensed and broad description of the phenomenon from analyzing written, verbal, or visual communication message (Elo & Kyngäs, 2008). The inductive approach begins by selecting a sample of text data without predefining the categories and guided by theories prior to the coding process. This can become challenging with large-scale social media data. Human coders are required to carefully read each piece of content with sufficient knowledge about the context to generate a category. This process may iterate multiple times, thus highly time-consuming and laborious. Recent advancements in LLMs offer promising alternatives to overcome these limitations. Many studies demonstrate that LLMs can match or exceed human performance in annotation tasks (Chew et al., 2023; Gilardi et al., 2023; Heseltine & Clemm von Hohenberg, 2024; Kuzman et al., 2023; Mu et al., 2024; Törnberg, 2024). Studies confirmed that LLMs, with the appropriate prompts, can be a strong alternative to traditional approaches of topic modeling in extracting topics (Mu et al., 2024). Some scholars are more conservative on the usage of LLMs. Their studies applied a human-LLM hybrid framework where the human expert works iteratively with LLMs to conduct thematic analysis and topic modeling (Dai et al., 2023; Heseltine & Clemm von Hohenberg, 2024). The inner-annotator agreement (IAA) of human and machine coder reached a Cohen's $\kappa$ equal to 0.81 (Dai et al., 2023).

Most of the research, however, remains technical. They are mostly conducted by Natural Language Processing (NLP) scholars focusing on zero-shot and few-shots prompt engineering and model refinement. The accuracy is evaluated by comparing model outputs to traditional machine learning benchmarks like BERT or other types of LLMs, usually via a case study (Azher et al., 2024; Kim et al., 2023; Stammbach et al., 2023). These efforts typically prioritize model improvement over real-world problems. Moreover, due to LLMs requiring large training sets, the analyses are focused on highly visible topics like the U.S. presidential election (Heseltine & Clemm von Hohenberg, 2024; Törnberg, 2024), COVID-19 (Mu et al., 2024), news articles (Kapoor et al., 2024), and public healthcare (Mirzaei et al., 2024). Few research applies LLM-supported thematic analysis to climate discourse. While sentiment analysis is a common approach (Chakraborty et al., 2020), some scholars argue sentiment analysis can only reveal surface-level attitudes on climate policies (Jost et al., 2019) or geospatial variations and polarizations (Dahal et al., 2019). Applying LLMs to conduct content analysis on climate change social media data can offer in-depth insights to the current literature.

## METHODS
### Dataset
YouTube users often prioritize conspiracy theories over scientific content (Allgaier, 2019; Bessi et al., 2016). Video content generally receives higher levels of engagements than text-based platforms like Twitter/X and Reddit (Yadav et al., 2011), making it ideal for studying ACC polarization. We collected comments from 30 YouTube videos using the YouTube API, selecting 15 videos supporting ACC ("believers") and 15 promoting climate hoax ("skeptics"), based on top-viewed results for the search terms "climate change" and "climate hoax". The dataset comprised 556,168 comments in total—326,065 from "believers" and 230,103 from "skeptics". As an exploratory study, we drew a 10% stratified sample for our final analysis.

### LLM Topic Annotation
LLMs offer scalable, cost-efficient alternatives to traditional annotation while maintaining interpretability and flexibility (Gilardi et al., 2023; Huang et al., 2023; Kuzman et al., 2023). We used GPT-4o-mini as a qualitative annotator to extract topics from user comments. Our two-step process was designed to generate interpretable and consistent topic labels while minimizing hallucination.

In Step 1, we selected a balanced sample of 500 comments by video type and video ID for prompt selection. While our prompt was inspired by (Mu et al., 2024; Stammbach et al., 2023) in extracting overarching topics, we added additional instructions asking the model to provide a rationale for each identified topic (see Figure 1). This rationale requirement was introduced to enhance the transparency of the model's decision-making and support validation in Step 2, where the topics would be applied to a larger dataset. These instructions were input as a system prompt, which sets the context and behavior expectations for the model during the task (Giray, 2023).

To evaluate the coherence of the extracted topics, we conducted two validation procedures. First, we manually reviewed the initial 500 comments to assess whether the model-generated topics and justifications effectively captured the content. Second, we tested generalizability by applying the model to a new sample of 500 comments. To account for variability in LLM responses, we repeated the annotation process three times and compared the Cohen's Kappa (Kim et al., 2023; Cohen, 1960). The score averaged .84, indicating high inter-run consistency and reliable topic assignment.

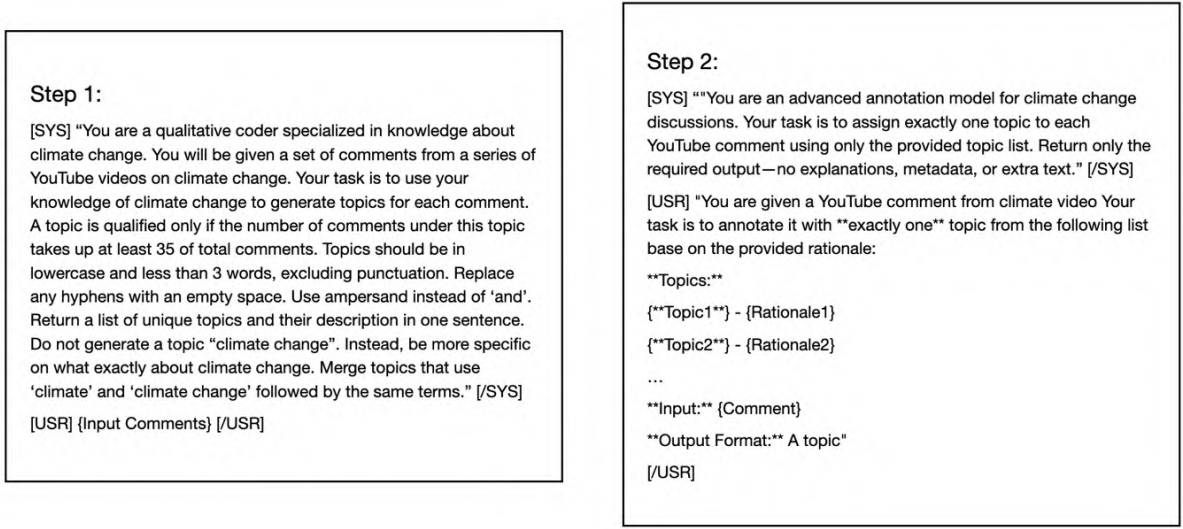

**Figure 1. Prompts used for Topic Annotation in step 1(left) and step 2(right).** [SYS] tag denotes system prompts, and [USR] tag user prompts. Curly brackets indicate variable input fields. Double asterisks (**) are used to indicate bold font effects within the prompt.

In Step 2, we applied the validated topics to a larger subset of 44,989 comments—roughly 10% of the total data. This subset included 19,992 from "believers" and 24,990 from "skeptics", balanced across video type and ID. Each comment was annotated using the previously generated topic list and accompanying rationales to guide consistent labeling. The specific prompt used in this step is also shown in Figure 1.

**Network**
Social network analysis (SNA) is applied to examine interaction patterns among commenters as it is an approach to study a set of socially relevant nodes connected by one or more relations (Scott & J.Carrington, 2014). We created two sets of *commenter-on* pairing edgelists for the change and hoax videos, respectively, where nodes represent users (change videos: N=7,798; hoax videos: N=8,015). The *commenter* is the user who commented and the *on* is the receiver of the comment, another user. Parent-child reply chains were fully unnested.

To visualize the networks, we used the *igraph* package in R. Two layout algorithms were employed: the Distributed Recursive Layout (DRL); and the Fruchterman-Reingold (FR) layout (Csardi & Nepusz, 2006). DRL is particularly well-suited for large and dense networks due to its scalable and iterative design, which efficiently positions nodes in two-dimensional space to reduce edge crossings and minimize node overlap. FR is particularly effective for small to moderate sized graphs. In the FR layout, densely connected nodes are drawn closer together, while sparsely connected nodes are pushed outward. We applied both layouts to each video and selected the most interpretable visualization. We also removed video ID nodes and their edges as we focused on the relationships between users and for clarity. Next, we merge the plot with the comments' topics data and colored nodes by topic (blue for users who commented to the topic, and salmon for others). Figure 2 shows a grid of 10 networks per video type, representing topic-specific engagement.

To answer our research question, we fit a Linear Mixed-Effects Regression Model (LMER) using normalized average node degree as the outcome. Fixed effects included topic and video type. Video ID was a random intercept to account for repeated measures across videos. We used restricted maximum likelihood estimation (REML) for model fitting and Satterthwaite's method for degrees of freedom and t-tests. The model showed a REML criterion of –568.2, and residuals were normally distributed with minor deviations in the upper tail. Variance inflation factors (VIFs) for the fixed effects were all well below critical thresholds (VIF < 2), indicating no collinearity concerns.

**RESULTS & DISCUSSION**
Our first step—prompting the model to output overarching topics along with justifications—resulted in 10 distinct topics, each grounded in a rationale provided by the LLM (see Table 1). The table lists these topics alongside concise explanations generated by the model, which clarify why each theme was identified as salient. This rationale helped ensure interpretability and provided an internal check on the coherence of topic assignments. During the broader annotation phase in Step 2, comments that did not align with any of the 10 predefined categories (n = 194) were flagged as outliers. These were manually reviewed by the authors and re-annotated based on their content and the model's accompanying rationale. Those that could not be assigned to one of the 10 topics were labeled as 'noise' (n = 184). The final distribution of topics for 44,989 comments is also described in Table 1.

| Topic | Rationale | Count |
|---|---|---|
| climate skepticism | Comments expressing doubt or denial regarding mainstream scientific consensus on climate change and its anthropogenic causes. | 17,256 |
| natural cycles | Perspectives emphasizing natural climatic cycles and historical data as evidence against the urgency attributed to human-induced climate change. | 6,589 |
| climate solutions | Discussions about proposed solutions to climate change issues, including renewable energy, reforestation, and technological innovations. | 5,183 |
| climate change misinformation | Discussions on misinformation related to climate change, including skeptics' claims and counterarguments against established scientific data. | 2,583 |
| greenhouse gases | Conversations centered on the role of greenhouse gases, particularly concerning human contributions and their impact on climate dynamics. | 2,544 |
| government policy | Discussions around how government policies, such as carbon taxes and environmental regulations, influence the conversation on climate change. | 2,505 |
| scientific consensus | References to the supposed agreement among scientists about climate change, often critiqued or supported within the comments. | 2,445 |
| economic impact | Insights into how climate policy and actions related to climate change affect economies, industries, and consumer behaviors. | 2,153 |
| environmental activism | Comments discussing the influence and role of environmental activism, including figures like Greta Thunberg and the narrative presented by activists. | 1,809 |
| media portrayal | Critiques regarding how media covers climate change issues and the perceived bias in the presentation of facts and interviews. | 1,731 |

**Table 1. Topics and rationale output from step 1 and final distribution of topics after step 2.**

By linking topic annotations to user interactions, Figure 2 visualizes topic-specific social networks for both video types. A-D represents four videos, 2 change and 2 hoax, selected by having the maximum and minimum number of nodes on topic 'climate change misinformation', which is the baseline of the regression model. 'Climate change misinformation' also has the largest variation in average degree, thus providing the most extreme visual contrast. We also highlighted 'government policy' and 'natural cycles.' Both topics show significant correlation in the model (see Table 2). Complementing the structural view of social networks, Figure 3 presents a box plot of the normalized average degree for each topic across video types. This visualization allows us to compare interaction intensity.

The overall number of 'climate skepticism' nodes is greatly more than the other types of topics in both climate change and hoax videos (see Figure 2). However, the average degree of these nodes is not contrastingly higher than the other topics (see Figure 3). This implies although many comments were mentioning skepticism on climate change, these comments do not necessarily receive much more attention from other users. Such result is comparatively more interesting for hoax videos. We suspect these comments are mostly subjective complaints expressing skepticism on climate change without having any (mis)information to defend their opinion. From the social network plot, topics such as 'climate solutions' and 'natural cycles' were mentioned frequently in both types of videos. Particularly, we notice that the average degree of nodes is relatively lower in hoax videos than change videos, implying that this is likely uncontested beliefs for people against ACC. We also noticed that the difference in total number of nodes in 'media portrayal' for both types of videos is not significant (N=341 for hoax videos and N=312 for change videos), whereas the average degree of hoax videos is substantially greater than that of change videos. For hoax videos this topic seems to be a more hot-button issue, likely tied to distrust in mainstream news. This results aligns with previous studies (Al-Rawi et al., 2021) that the notion of "fake news" is being contested and weaponized by various ideologically-motivated actors to attack and discredit scientific consensus on ACC.

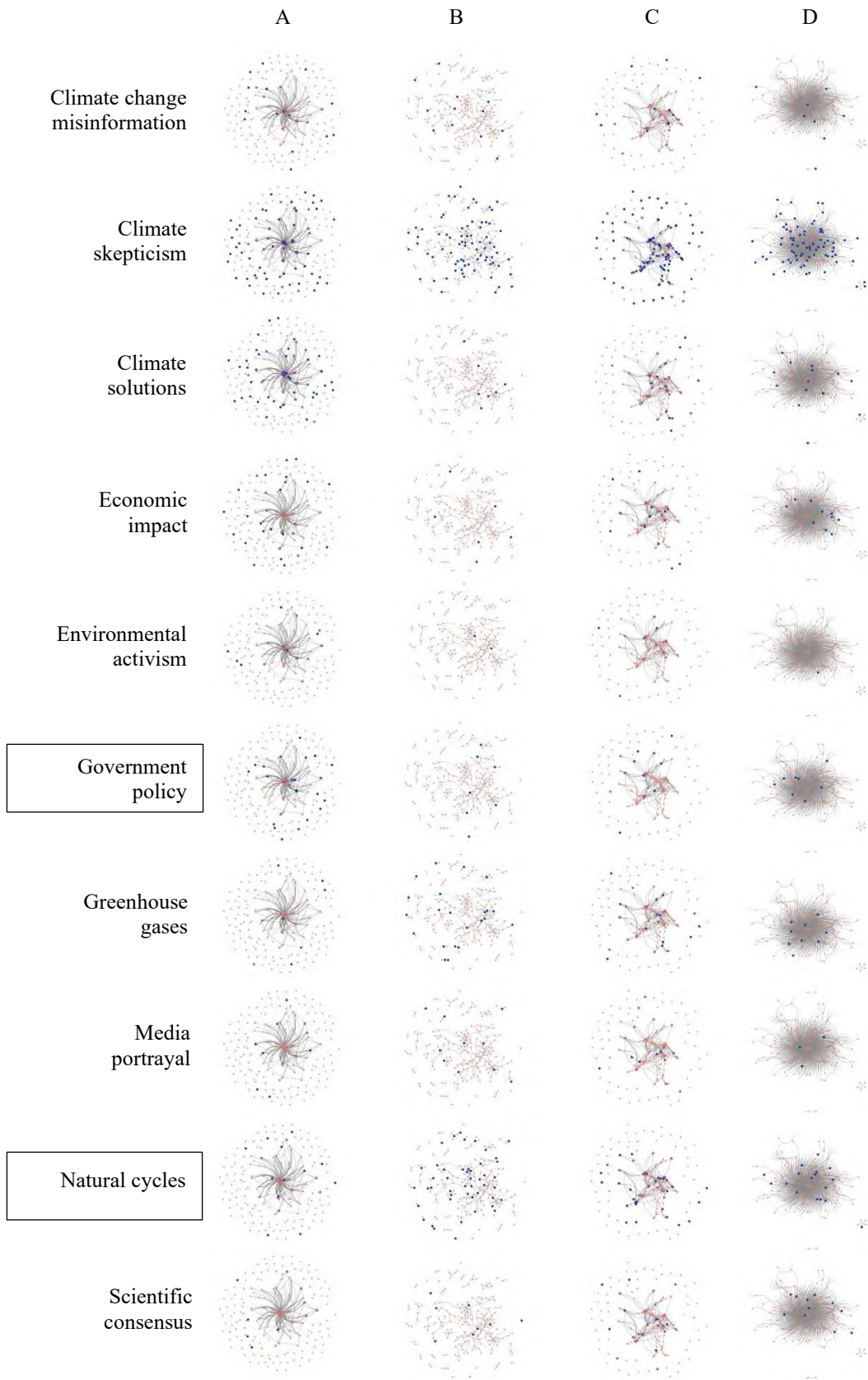

**Figure 2. Social network plots by topics** (A and B: change videos with maximum and minimum number of nodes in topic 'climate change misinformation'. C and D: hoax videos with maximum and minimum number of nodes in topic 'climate change misinformation'.) A and C are plotted in DRL layout, B and D are plotted in FR layout.

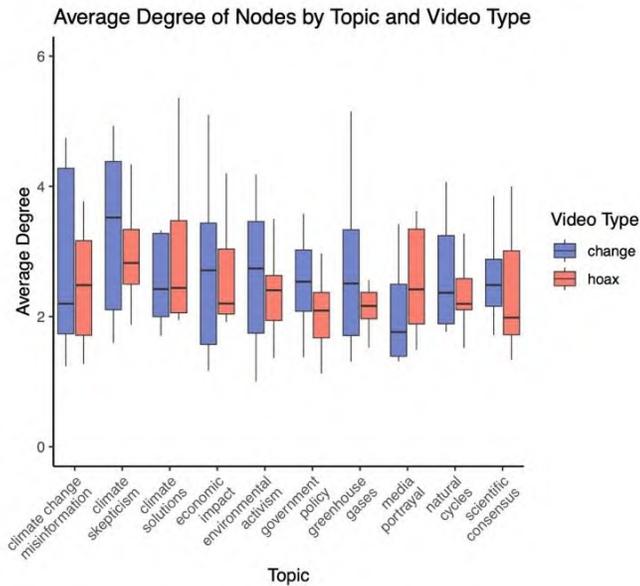

**Figure 3. Average Degree of Nodes by Topic and Video Type**

In answering our research question—how climate change "believers" and "skeptics" differ in their YouTube comment discourse—our model made a prediction with 'climate change misinformation' as the reference topic and 'hoax' as the baseline video type. Table 2 summarizes the model estimates. The results indicate that two topics were significantly associated with lower normalized average degree: 'government policy' ($β$ = -0.039, $p$ = .039) and 'natural cycles' ($β$ = -0.038, $p$ = .048). This suggests that comments on these topics were less likely to generate replies or interaction from other users. Interestingly, no significant difference was found between the video types (change vs. hoax), indicating that video stance alone does not significantly predict user engagement once topic is accounted for ($β$ = -0.013, $p$ = .356).

The relatively lower engagement with government policy may reflect a pattern where users treat policy-oriented statements as declarative or non-controversial within their ideological community (Kahan et al., 2011; McCright & Dunlap, 2011; Nisbet, 2009). For instance, believers may voice support for regulation, while skeptics reject climate policy as government overreach—yet in both cases, such statements may affirm rather than invite further discussion. Similarly, the natural cycles topic tends to be a conversational endpoint for skeptics as it is often used to deny anthropogenic causes. Believers, on the other hand may be unwilling to engage with such claim that they perceive as scientifically inaccurate or that challenges their group's position (Kahan, 2013).

Notably, 'media portrayal' was not significantly associated with higher or lower interaction ($p$ = .350), despite the average degree of nodes being high in hoax videos. This may imply that although the media is a frequent target of discourse, it does not necessarily drive extended conversation. While "fake news" narratives are common in climate denial discourse, they are often used as identity-affirming signals rather than debate starters (Pearce et al., 2014; Zollo, 2019).

Surprisingly, video type was not a significant predictor of engagement, suggesting that the stance of a video does not, in itself, lead to more or less interaction among commenters. Instead, it is the comments content, or topic, that shapes how discussions unfold. This finding reinforces the idea that motivated reasoning operates at the discourse level (Druckman & McGrath, 2019), influencing what users choose to talk about and how they respond to others.

These findings are consistent with recent shifts in the nature of online climate denial discourse. As (de Nadal, 2024) notes, climate skeptics on YouTube increasingly frame their resistance not as scientific disagreement but as opposition to perceived overreach by activists, elites, or the government—aligning closely with culture war narratives. This helps explain why "climate change misinformation" elicits higher engagement, while topics like "government policy" or "natural cycles" act as ideological endpoints within belief communities, receiving fewer replies. Such patterns reinforce the idea that content perceived as controversial or identity-threatening provokes more interaction, whereas content that affirms group consensus may be met with silence.

The absence of a significant video-type effect also aligns with broader studies showing that climate discourse polarization is shaped more by identity and network dynamics than content framing alone (Bliuc et al., 2015; Falkenberg et al., 2022; Yantseva, 2024). In highly polarized environments, individuals often use climate discourse to signal membership in ideological groups rather than engage in deliberation. Our network analysis reflects this: even when discussing similar topics, believers and skeptics do so in distinct and insular ways. These results support recent reviews (Sultana et al., 2024; van Eck, 2024), who highlight emotionally charged and segregated discourse as key characteristics of climate communication online.

Overall, the results help explain our research question on what motivates people to join in discourses regarding climate change in YouTube videos. Topics that are ideologically settled within belief groups, such as 'natural cycles' and 'government policy', serve as echo chamber statements (Zollo, 2019), with low engagement between users. On the other hand, considering that other topics have a negative average degree compared to 'climate change misinformation', it suggests that misinformation related comments may function as provoking discourse, prompting more responses and interactions from users. The absence of a significant effect for what the video ultimately is about, further emphasizes that interactional dynamics are shaped more by content than context, showing how motivated reasoning process manifests at the conversational level. Our research illustrates how LLM-supported qualitative analysis uncovers not just what people debate with respect to climate change, but how the discourse is motivationally structured.

| Variable | Estimate | Std. Error | df | t value | Pr(>|t|) | Significance |
|---|---|---|---|---|---|---|
| (Intercept) | 0.05712 | 0.01584 | 142.38 | 3.606 | 0.000429 | *** |
| climate skepticism | -0.03264 | 0.01901 | 225.00 | -1.717 | 0.087403 | . |
| environmental activism | -0.03664 | 0.01901 | 225.00 | -1.927 | 0.055183 | . |
| government policy | -0.03948 | 0.01901 | 225.00 | -2.077 | 0.038974 | * |
| economic impact | -0.03704 | 0.01901 | 225.00 | -1.948 | 0.052633 | . |
| climate solutions | -0.03451 | 0.01901 | 225.00 | -1.815 | 0.070827 | . |
| greenhouse gases | -0.03493 | 0.01901 | 225.00 | -1.837 | 0.06749 | . |
| media portrayal | -0.01765 | 0.01901 | 225.00 | -0.928 | 0.354295 | |
| natural cycles | -0.03773 | 0.01901 | 225.00 | -1.985 | 0.048384 | * |
| scientific consensus | -0.03729 | 0.01901 | 225.00 | -1.961 | 0.051077 | . |
| change videos | -0.01301 | 0.01383 | 24.00 | -0.941 | 0.356113 | |

**Table 2. Linear mixed-effect model results** (0: '***'; 0.001: '**'; 0.01: '*'; 0.05: '.'; 0.1: ' ')

## FUTURE DIRECTIONS

While our study highlights the potential of LLMs in scaling qualitative content analysis, some areas remain open for further exploration. We analyzed a stratified 10% sample of the full dataset, carefully balanced by video type and ID to ensure representativeness. Although this approach enabled in-depth analysis at a manageable scale, future research could extend this work to the full dataset or other platforms to assess broader generalizability. Still, our findings provide a compelling view into the content and structure of climate discourse on YouTube, and how such conversations reflect underlying motivations.

In addition, our research lacks the comparison between the model's annotation and human-coded labels, which is a common benchmark for evaluating classification accuracy. However, we incorporated rationale-based prompting, repeated the annotation process across three trials to assess consistency, and manually reviewed whether the extracted topics adequately captured overarching themes in the data. These steps helped ensure both interpretability and internal validity, and future work could build on this by including human-coded evaluations or comparing model performance across different LLMs and architectures such as BERT.

## CONCLUSION

This study explored how climate change "believers" and "skeptics" differ in their YouTube comment discourse, focusing on the topics they emphasize and the interaction with other users. Annotating topics using LLM and incorporating social network analysis, we found that interaction patterns are driven more by the content of the comment than the video stance. Topics such as government policy and natural cycles tend to reinforce group consensus with limited engagement, while climate change misinformation generates more interaction, likely due to its provocative or emotionally charged nature.

These patterns reflect motivated reasoning, where users engage selectively with content aligned with their identity and ideology. Our approach demonstrates how LLMs can scale qualitative analysis and help reveal how belief systems shape online climate discourse. Future research may build on this work to design more inclusive and productive strategies for climate communication online.

## GENERATIVE AI USE

We employed ChatGPT (4o-mini model) for the following purposes: Content annotation and surfacing additional climate change literature. We evaluated the output of the annotation by cross-validating through 3 different run sessions and manually examining the output (see Methodology section for details). The authors assume all responsibility for the content of this submission.

## AUTHOR ATTRIBUTION

Yuheun Kim: conceptualization, methodology, data curation, formal analysis, writing–original draft. Qiaoyi Liu: conceptualization, methodology, data curation, formal analysis, visualization, writing–original draft. Jeff Hemsley: project administration, conceptualization, supervision, writing-review and editing